\documentclass[aps,prl,twocolumn,superscriptaddress]{revtex4-1}
\usepackage{amsmath,amsfonts,amssymb}
\usepackage{graphicx,float,calc}
\usepackage{color,bm}
\usepackage{ulem}
\usepackage{braket}
\usepackage[colorlinks,urlcolor=blue,citecolor=blue,linkcolor=blue]{hyperref}

\begin{document}

\title{Connecting Topological Anderson and Mott Insulators in Disordered Interacting Fermionic Systems}

\author{Guo-Qing Zhang}
\affiliation{Guangdong Provincial Key Laboratory of Quantum Engineering and Quantum Materials, School of Physics and Telecommunication Engineering, South China Normal University, Guangzhou 510006, China}
\affiliation{Guangdong-Hong Kong Joint Laboratory of Quantum Matter, Frontier Research Institute for Physics, South China Normal University, Guangzhou 510006, China}
\author{Ling-Zhi Tang}
\affiliation{Guangdong Provincial Key Laboratory of Quantum Engineering and Quantum Materials, School of Physics and Telecommunication Engineering, South China Normal University, Guangzhou 510006, China}
\author{Ling-Feng Zhang}
\affiliation{Guangdong Provincial Key Laboratory of Quantum Engineering and Quantum Materials, School of Physics and Telecommunication Engineering, South China Normal University, Guangzhou 510006, China}
\author{Dan-Wei Zhang}\thanks{danweizhang@m.scnu.edu.cn}
\affiliation{Guangdong Provincial Key Laboratory of Quantum Engineering and Quantum Materials, School of Physics and Telecommunication Engineering, South China Normal University, Guangzhou 510006, China}
\affiliation{Guangdong-Hong Kong Joint Laboratory of Quantum Matter, Frontier Research Institute for Physics, South China Normal University, Guangzhou 510006, China}
\author{Shi-Liang Zhu}
\affiliation{Guangdong Provincial Key Laboratory of Quantum Engineering and Quantum Materials, School of Physics and Telecommunication Engineering, South China Normal University, Guangzhou 510006, China}
\affiliation{Guangdong-Hong Kong Joint Laboratory of Quantum Matter, Frontier Research Institute for Physics, South China Normal University, Guangzhou 510006, China}

\begin{abstract}
The topological Anderson and Mott insulators are two phases that have so far been separately and widely explored beyond topological band insulators. Here we combine the two seemingly different topological phases into a system of spin-1/2 interacting fermionic atoms in a disordered optical lattice. We find that the topological Anderson and Mott insulators in the noninteracting and clean limits can be adiabatically connected without gap closing in the phase diagram of our model. Lying between the two phases, we uncover a disordered correlated topological insulator, which is induced from a trivial band insulator by the combination of disorder and interaction, as the generalization of topological Anderson insulators to the many-body interacting regime. The phase diagram is determined by computing various topological properties and confirmed by unsupervised and automated machine learning. We develop an approach to provide a unified and clear description of topological phase transitions driven by interaction and disorder. The topological phases can be detected from disorder/interaction induced edge excitations and charge pumping in optical lattices.

\end{abstract}

\date{\today}

\maketitle

{\color{blue}\textit{Introduction.---}}Topological insulators with intriguing properties that are robust against certain disorders or interactions have been well understood in the framework of topological band theory \cite{Hasan2010,XLQi2011,Bansil2016}. For strong disorder or interaction, the energy bands generally become trivial due to the Anderson or Mott localization \cite{Anderson1958,Belitz1994}. Strikingly, it was found that a topological phase can be induced from a trivial phase by disorders, known as topological Anderson insulator (TAI) \cite{JLi2009}. The TAIs have been theoretically studied \cite{JLi2009,Groth2009,HJiang2009,CZChen2015,HMGuo2010,Altland2014,Mondragon-Shem2014,Titum2015,Sriluckshmy2018,RChen2019,DWZhang2020,XWLuo2019,WHong2020,XSWang2020,CLi2020,YBYang2021} and experimentally observed in various system, such as ultracold atoms \cite{Meier2018a}, photonic and sonic crystals \cite{Stutzer2018,GGLiu2020, Zangeneh-Nejad2020}, and electric circuits \cite{WZhang2021}. On the other hand, interactions can induce correlated topological phases \cite{Rachel2018}, such as fractional quantum Hall effects \cite{Tsui1982} and topological Mott insulator (TMI) \cite{Raghu2008}. TMIs are a class of interaction-induced topological phases for interacting fermions or bosons \cite{Raghu2008,YZhang2009,Pesin2010,Yoshida2014,Herbut2014,Amaricci2015,Imriska2016,Barbarino2019,Irsigler2019,Dauphin2012,SLZhu2013,XDeng2014,Kuno2017,Grusdt2013,ZXu2013,HHu2019,YLChen2020,DWZhang2020b,YXWang2019}, characterized by many-body topological invariants and edge excitations. %andrealizable with cold atoms \cite{Dauphin2012,SLZhu2013,XDeng2014,Kuno2017,Grusdt2013,ZXu2013,HHu2019,YLChen2020}.
Despite the disorder or interaction induced topological (localization) phenomena are of growing interest, a unified picture of these two routes has not emerged so far. Remarkably, it remains unclear whether the TMI and TAI can exhibit in a system, as these two seemingly different topological phases are separately proposed. If yes, how can they be connected and what are the phases lying between them? Can TAIs exhibit within interactions in view of current studies only focusing on the single-particle TAIs? These questions involve the interplay among interaction, disorder, and topology, which are fundamentally important but challenging in condensed matter physics and artificial quantum systems.

%It remains unclear whether the interaction-induced TMI and disorder-induced TAI can exhibit in a system, as these two seemingly different topological phases are separately explored so far. If yes, how can they be connected and what are the phases lying between them? Can the TAIs exhibit in the presence of interactions in view of the studies of TAIs up to now only focusing on the
%single-particle regime? These questions are particularly interesting and important to explore quantum states of matter from the interplay among topology, disorder, and interaction.

In this Letter, we address these questions by exploring topology of spin-1/2 interacting fermions in a disordered 1D optical lattice (OL). Our main results are: (i) We combine the TAI and TMI in the noninteracting and clean limits into this system and find that they can be adiabatically connected without gap closing. (ii) Lying between them, we uncover a disordered correlated topological insulator (DCTI) induced by the combination of disorder and interaction from a trivial band insulator (BI), as the first generalization of noninteracting TAIs to the many-body interacting regime. (iii) We not only numerically compute various topological properties and use unsupervised and automated machine learning to determine the phase diagram, but also develop an analytical unified approach to reveal topological phase transitions driven by the interaction and disorder with the mean-field (MF) and self-consistent Born approximations (SCBAs). The revealed topological phases can be detected from the interaction/disorder induced topological edge excitations and charge pumping in OLs. %Our work paves the way to search on exotic topological phases in disordered interacting systems.

\begin{figure}[t]
\centering
\includegraphics[width=0.48\textwidth]{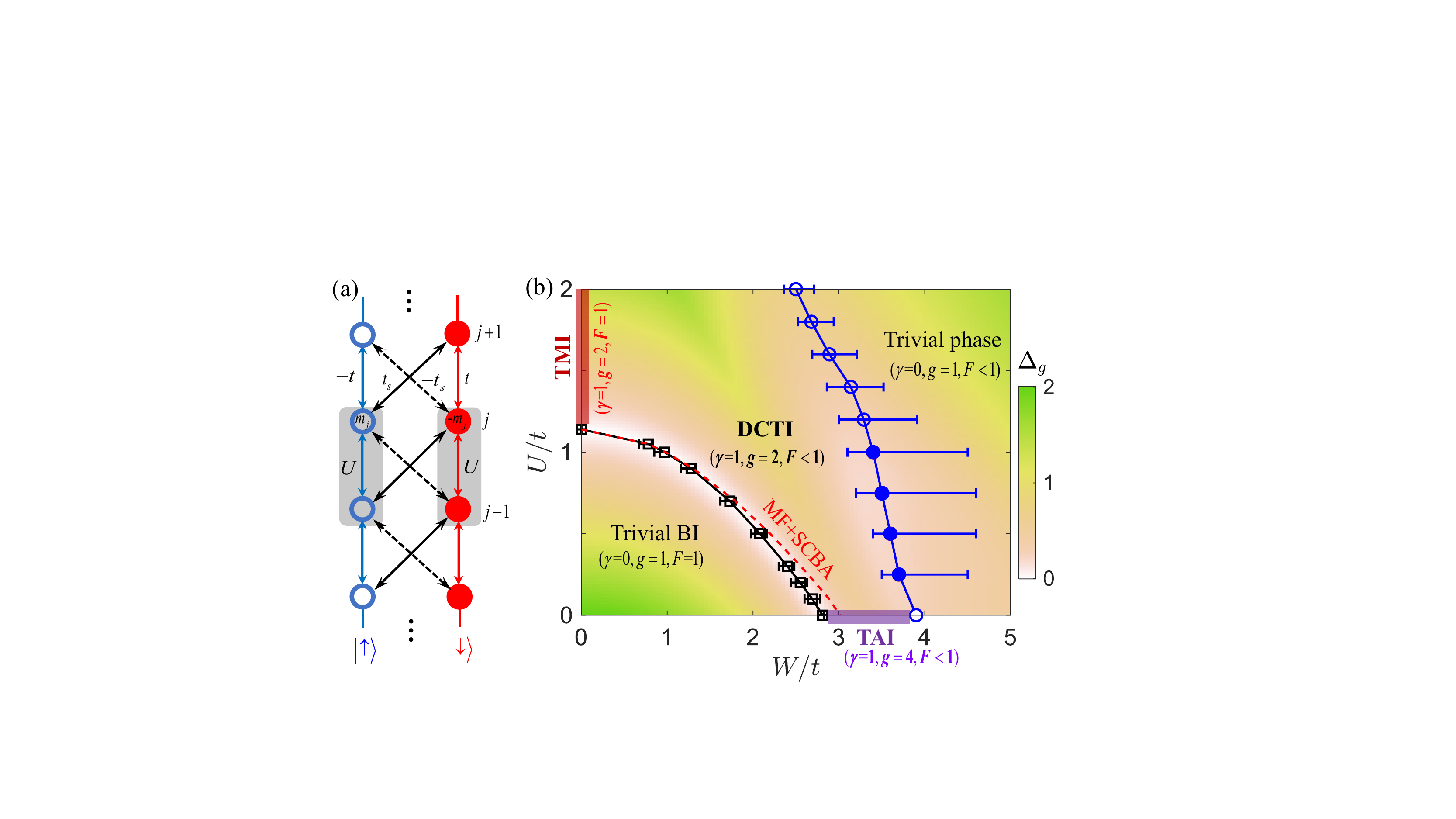}
\caption{(Color online) (a) Proposed model for interacting fermions in a spin-dependent optical ladder with disordered potentials. (b) Phase diagram of the Hamiltonian (\ref{Ham}) on the $W$-$U$ plane. The square and circle symbols guided by the solid lines are determined from the Berry phase, and the error bars denotes the uncertainty due to disorders. The TMI and TAI phases in the clean and noninteracting limits are connected via the DCTI phase and become a disordered trivial phase for stronger disorders. The red dashed line denotes the phase boundary between the trivial BI and the topological phases, obtained using the MF and SCBA approaches. The excitation gap $\Delta_g$ is plotted as the background. Parameters are $t=1$, $t_{s}=0.95$, $m_z=3$, $\alpha=(\sqrt{5}-1)/2$, $L=N_a=24$ from the DMRG for four solid blue circles, otherwise $L=N_a=8$ for $U\neq0$ and $L=610$ for $U=0$ from the ED.}
\label{fig1}
\end{figure}

{\color{blue}\textit{Model and phase diagram.---}}We consider an atomic gas of spin-1/2 interacting fermions in a 1D spin-dependent OL with disordered ladder potentials \cite{Creutz1999,Atala2014,Junemann2017,JHKang2018,Roati2008,Schreiber2015,XLi2013,XZhou2017,Goldman2014,DWZhang2018,Cooper2019}, as shown in Fig. \ref{fig1}(a). The tight-binding Hamiltonian reads
\begin{equation}\begin{split}\label{Ham}
H=&-t\sum_{j}(c_{j \uparrow}^{\dagger} c_{j+1 \uparrow}-c_{j \downarrow}^{\dagger} c_{j+1 \downarrow}+\mathrm{H.c.})\\
&+t_{s}\sum_{j}(c_{j \uparrow}^{\dagger} c_{j+1 \downarrow}-c_{j \downarrow}^{\dagger} c_{j+1 \uparrow}+\mathrm{H.c.})\\
&+\sum_{j} m_{j}(n_{j \uparrow}-n_{j \downarrow})+U\sum_{j\sigma=\uparrow,\downarrow}n_{j\sigma} n_{j+1 \sigma},
%\hat{H}=&-t\sum_{j}(\hat{c}_{j \uparrow}^{\dagger} \hat{c}_{j+1 \uparrow}-\hat{c}_{j \downarrow}^{\dagger} \hat{c}_{j+1 \downarrow}+\mathrm{H.c.})\\
%&+t_{s}\sum_{j}(\hat{c}_{j \uparrow}^{\dagger} \hat{c}_{j+1 \downarrow}-\hat{c}_{j \downarrow}^{\dagger} \hat{c}_{j+1 \uparrow}+\mathrm{H.c.})\\
%&+\sum_{j} m_{j}(\hat{n}_{j \uparrow}-\hat{n}_{j \downarrow})+U\sum_{j\sigma=\uparrow,\downarrow}\hat{n}_{j\sigma} \hat{n}_{j+1 \sigma},
\end{split}\end{equation}
where $n_{j \sigma}=c_{j \sigma}^{\dagger}c_{j \sigma}$ and $c_{j \sigma}^{\dagger}$ ($c_{j \sigma}$) creates (annihilates) a spin-$\sigma$ fermion ($\sigma=\uparrow,\downarrow$) at site $j$ ($j=1,2,...,L$) of the two-leg ladder of length $L$. Here $t$ ($t_{s}$) denotes the spin-dependent (spin-flip) hopping and $t\equiv1$ as the energy unit, $m_{j}$ is the lattice potential as an effective diordered Zeeman field, and $U$ is the inter-leg (spin) interaction (see the intra-leg interaction in the SM \cite{Note1}). We focus on the half-filling case with atomic number $N_a=L$ and the quasiperiodic disorder $m_{j}=m_z+W\cos(2\pi\alpha j+\varphi)$ with $m_z$ the overall Zeeman strength, $W$ the disorder strength, $\alpha=(\sqrt{5}-1)/2$ an irrational number, and $\varphi$ an offset phase randomly chosen for sampling. The spin-dependent OL can be realized using proper laser beams with incommensurate potential for two atomic internal states with opposite detunings \cite{Atala2014,Junemann2017,JHKang2018,Roati2008,Schreiber2015,YWang2020}. The required spin-flip and spin-dependent hoppings have been proposed \cite{XJLiu2013} and experimentally realized \cite{ZWu2016,BSong2018} by Raman-assisted tunnelings in OLs. The interaction terms can be engineered for magnetic atoms with long-range dipolar interactions \cite{Trefzger2011,Dutta2015,Xu2013}. By aligning the dipolar atoms at the magic angle \cite{Trefzger2011}, only the inter-leg interaction is relevant (with vanishing intra-leg interactions). The cold-atom scheme to realize various tunable hopping and interaction terms in optical ladders was proposed in Ref. \cite{Junemann2017}.

The model in the $U=W=0$ limit has a topological insulator protected by the chiral symmetry in the AIII class for $|m_z|<2t$ \cite{XJLiu2013,Note1}. The disorder and interaction considered here preserve the chiral symmetry. We study the many-body ground state of the Hamiltonian (\ref{Ham}) by using the exact diagonalization (ED) and density-matrix renormalization  group (DMRG) methods \cite{Weinberg2019,White1992,Schollwoeck2011}, the unsupervised and automated machine learning \cite{vanNieuwenburg2017,Rodriguez-Nieva2019,PhysRevLett.125.170603,Kaeming2021}, and the MF and SCBA approaches \cite{Groth2009,HJiang2009,CZChen2015}. The quantities are averaged over 20 quasiperiodic configurations in numerical simulations. The random disorder case with similar results is given in the Supplemental Material (SM) \cite{Note1}.

Our main results are summarized in the ground-state phase diagram on the $W$-$U$ plane [Fig. \ref{fig1}(b)]. The ordered and disordered phases are described by the unity and non-unity fidelity [$\mathcal{F}$ in Eq. (\ref{fidelity})], respectively. The topology is characterized by the Berry phase [$\gamma$ in Eqs. (\ref{BP1},\ref{BP2})]. For small $W$ and $U$, the ground state is an ordered trivial BI. In the clean (noninteracting) limit $W=0$ ($U=0$), the ground state becomes an ordered TMI (disordered TAI) with increasing $U$ ($W$) up to the topological phase transition point $U_c\approx 1.14$ ($W_{c}\approx 2.8$). For finite $U$ and $W$, a DCTI phase with disordered ground states exhibits. Strikingly, the DCTI can be driven from the trivial BI with increasing $W$ for $U\lesssim1.14$ with a topological phase transition. The first phase boundary between the trivial BI and topological phases are obtained by computing the Berry phase and complementary quantities, which agrees well with the analytical results by the MF and SCBA approaches. For strong disorder, the DCTI and TAI become a disordered trivial phase with the second phase boundary. The gap $\Delta_g=E_1-E_0$ between the ground and first excited states with energies $E_{0,1}$ under periodic boundary condition (PBC) is plotted as the background of Fig. \ref{fig1}(b). We find the bulk gap closing (unclosing) in the first (second) phase boundary. Moreover, the TAI and TMI are adiabatically connected via the DCTI without gap closing in the phase diagram. The DCTI can be viewed as an intermediate phase between the TMI and TAI at the mean-field level, which behaves as a disordered TMI when $U\gtrsim1.14$ and a correlated TAI in the other region that broadens the concept of TAIs to the many-body interaction regime. The TAI, TMI and DCTI share similar topological characters (quantized Berry phase and edge modes) and compose a unified and clear description of disorder and interaction induced topological phases. Below we elaborate these topological phases.

Before proceeding, we make some remarks on the phase diagram. i) The finite-size scalings of the Berry phase and bulk gap are used to confirm the topological phases and their adiabatic connections, based on the DMRG for the system up to $L=N_a=48$ \cite{Note1}. ii) The DCTI and TMI belong to the symmetry-protected topological phases of interacting fermions \cite{XChen2012,CWang2014,Morimoto2015}. Their ground states under OBC near half filling are twofold degenerate with each collective mode occupied on each edge (see Fig. \ref{fig4}). Here the ground state degeneracy $g$ is refereed to counting the degeneracy among energies $\{E_0^{L},E_1^{L},E_0^{L\mp1}\}$, where $E_{0,1}^{N_a}$ denote the two lowest energies for $N_a=L-1,L,L+1$ under OBC. For the TMI and DCTI with $U>0$, one has $g=2$ as $E_0^{L}=E_1^{L}\neq E_0^{L\mp1}$ corresponding two collective edge excitations. For noninteracting TAIs (trivial phases), the degeneracy tends to $g=4$ ($g=1$ without edge modes) as $E_0^{L}=E_1^{L}=E_0^{L\mp1}$, related to two zero-energy single-particle edge modes either be empty or occupied. The subtle difference of edge modes in interacting and noninteracting topological phases can be observed \cite{Leseleuc2019,Note1}. iii) We use the machine learning approach in a unsupervised and automated fashion \cite{vanNieuwenburg2017,Rodriguez-Nieva2019,PhysRevLett.125.170603} to confirm the phase diagram with the well matched phase boundaries \cite{Note1}. iv) The DCTI driven by disorder and interaction can exhibit while the TAI is absent for certain values of $m_z$ \cite{Note1}. v) Our main results preserve even if an intra-leg interaction term $V\sum_jn_{j\uparrow}n_{j\downarrow}$ is tuned on \cite{Note1}.

%First, it is crossover instead of phase transition between the disordered TMI and correlated TAI, which correspond to the Mott-like and Anderson-like insulators, respectively. The MI is rigorously defined only for $W=0$ and the AI only for $U=0$, but one may refer to disordered MIs and correlated AIs in the presence of nonzero $W$ and $U$ \cite{Byczuk2005,Byczuk2010,TMa2018}. We roundly determine the two phases numerically using the MF energy variance $\Delta E_{\text{MF}}$ for the lowest MI in the clean limit of $U=U_c$ and $W=0$ (see the SM \cite{Note1}).

\begin{figure}[t]
\centering
\includegraphics[width=0.48\textwidth]{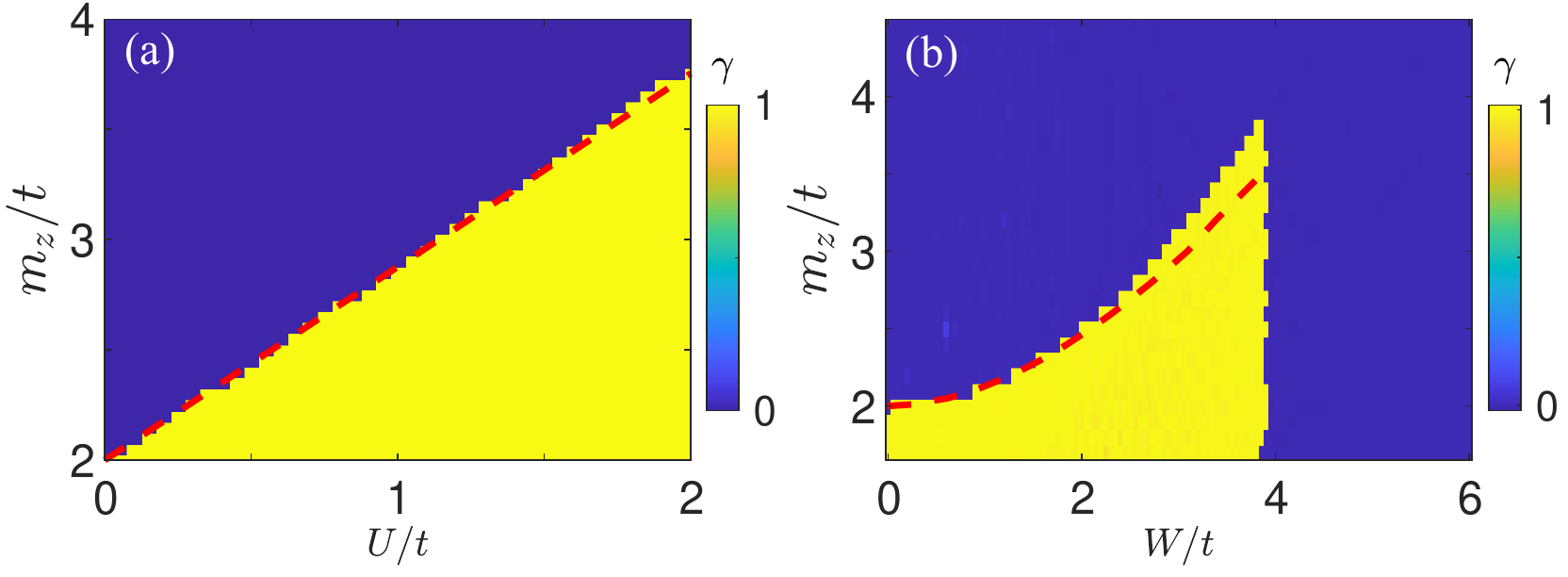}
\caption{(Color online) Berry phase $\gamma$ obtained from the ED as functions of $U$ and $m_z$ for $W=0$ and $L=N_a=8$ (a); and $W$ and $m_z$ for $U=0$ and $L=610$ (b), with the same other parameters as in Fig. \ref{fig1}. The red dashed lines are the phase boundaries obtained by the MF and SCBA approaches.}
\label{fig2}
\end{figure}

{\color{blue}\textit{TMI and TAI in two limits.---}} The topology of the system can be characterized by the Berry phase quantized in units of $\pi$ under the twisted PBC \cite{HGuo2011,Junemann2017}:
\begin{equation}\label{BP1}
\gamma=\frac{1}{\pi}\oint d \theta\bra{\Psi^g(\theta)}i \partial_{\theta}\ket{\Psi^g(\theta)} ~\text{mod} ~2,
\end{equation}
where $\ket{\Psi^g(\theta)}$ is the many-body ground state at half filling with the twist boundary phase $\theta$ \cite{Xiao2010}. Here $\gamma=1$ and 0 for topological and trivial phases, respectively. In the clean limit $W=0$, we compute $\gamma$ and obtain the topological phase diagram on the $U$-$m_z$ plane, as shown in Fig. \ref{fig2}(a). The ground state lies at the trivial BI phase for $m_z>2$ and $U=0$, and becomes a TMI with increasing $U$ up to the critical interaction $U_c$. We find that $U_c$ linearly depends on $m_z$ and $U_c\approx1.14$ for $m_z=3$. We also compute $\Delta_g$ to confirm the topological phase transition accompanied with gap closing and reopening.

In the $U=0$ limit, the Berry phase is rewritten as
\begin{equation}\label{BP2}
\gamma=\sum_{l=1}^{L} \frac{1}{\pi} \oint d\theta \bra{\psi_l(\theta)} i\partial_{\theta} \ket{\psi_l(\theta)} ~\text{mod} ~2,
\end{equation}
where $|\psi_l\rangle$ is the $l$-th single-particle wave function by the ED under the twisted PBC. Alternatively, we compute the winding number to reveal the same topology \cite{Note1}. The numerical result of $\gamma$ on the $W$-$m_z$ plane is shown in Fig.~\ref{fig2}(b), which reveals the disorder-induced TAI phase when $2\lesssim m_z\lesssim 3.8$. For $m_z=3$ in Fig. \ref{fig1}(b), we obtain the TAI for $2.8\lesssim W\lesssim3.9$. Note that the single-particle energy gap also closes at the topological phase transition from the trivial BI to the TAI.

\begin{figure}[t]
\centering
\includegraphics[width=0.48\textwidth]{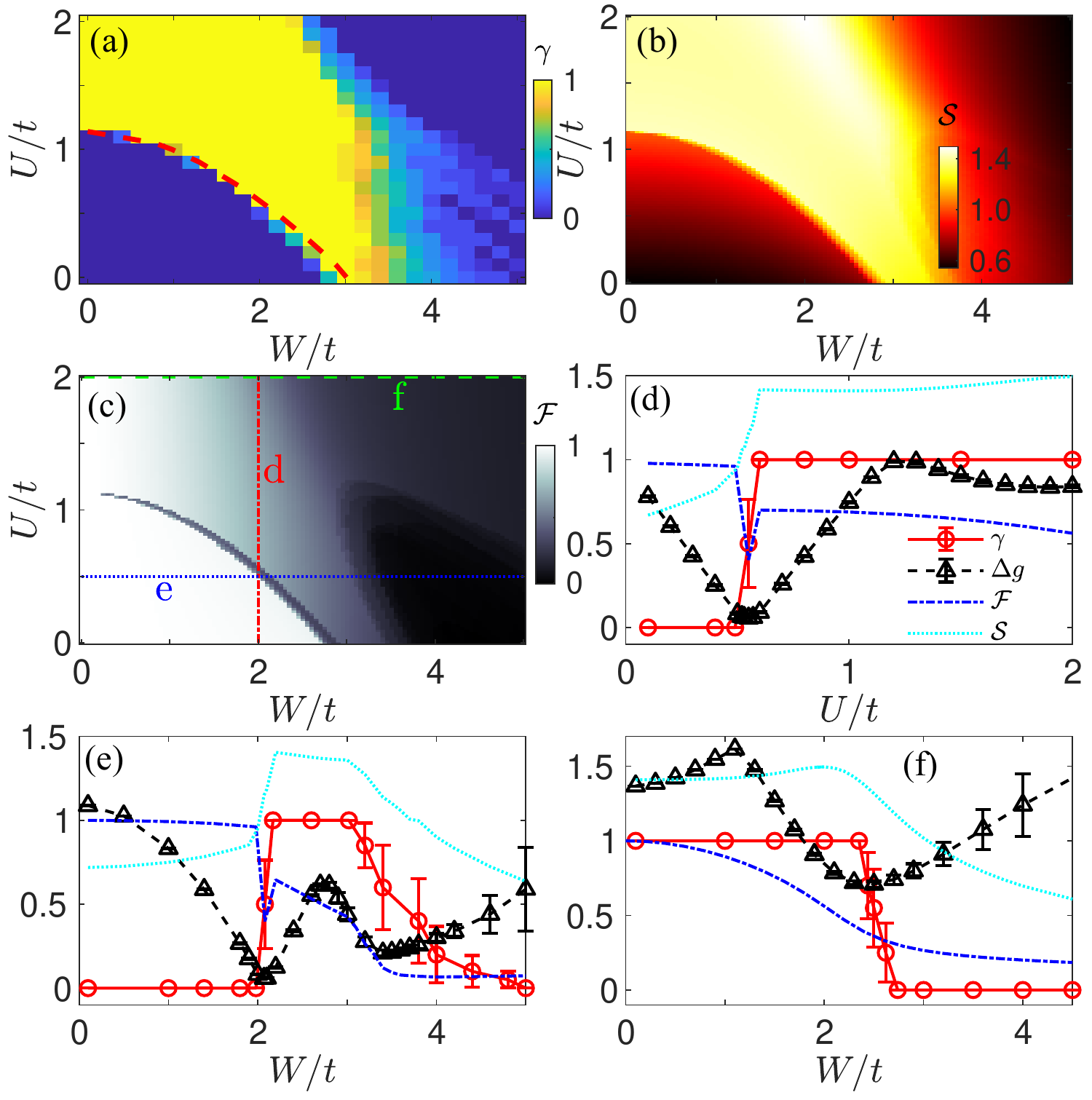}
\caption{(Color online) (a) Berry phase $\gamma$, (b) half-chain entanglement $\mathcal{S}$, and (c) ground-state fidelity $\mathcal{F}$ as functions of $W$ and $U$. The red dashed line in (a) is the phase boundary under the MF and SCBA approaches. (d-f) The quantities $\gamma$, $\Delta_g$, $\mathcal{F}$ and $\mathcal{S}$ as functions of $U$ or $W$ for $W=2$ (d), $U=0.5$ (e), and $U=2$ (f), as indicated by three cutting lines in (c). The results are obtained from the ED for $L=N_a=8$ and other parameters the same as in Fig. \ref{fig1}.}
\label{fig3}
\end{figure}

\begin{figure}[t]
\centering
\includegraphics[width=0.48\textwidth]{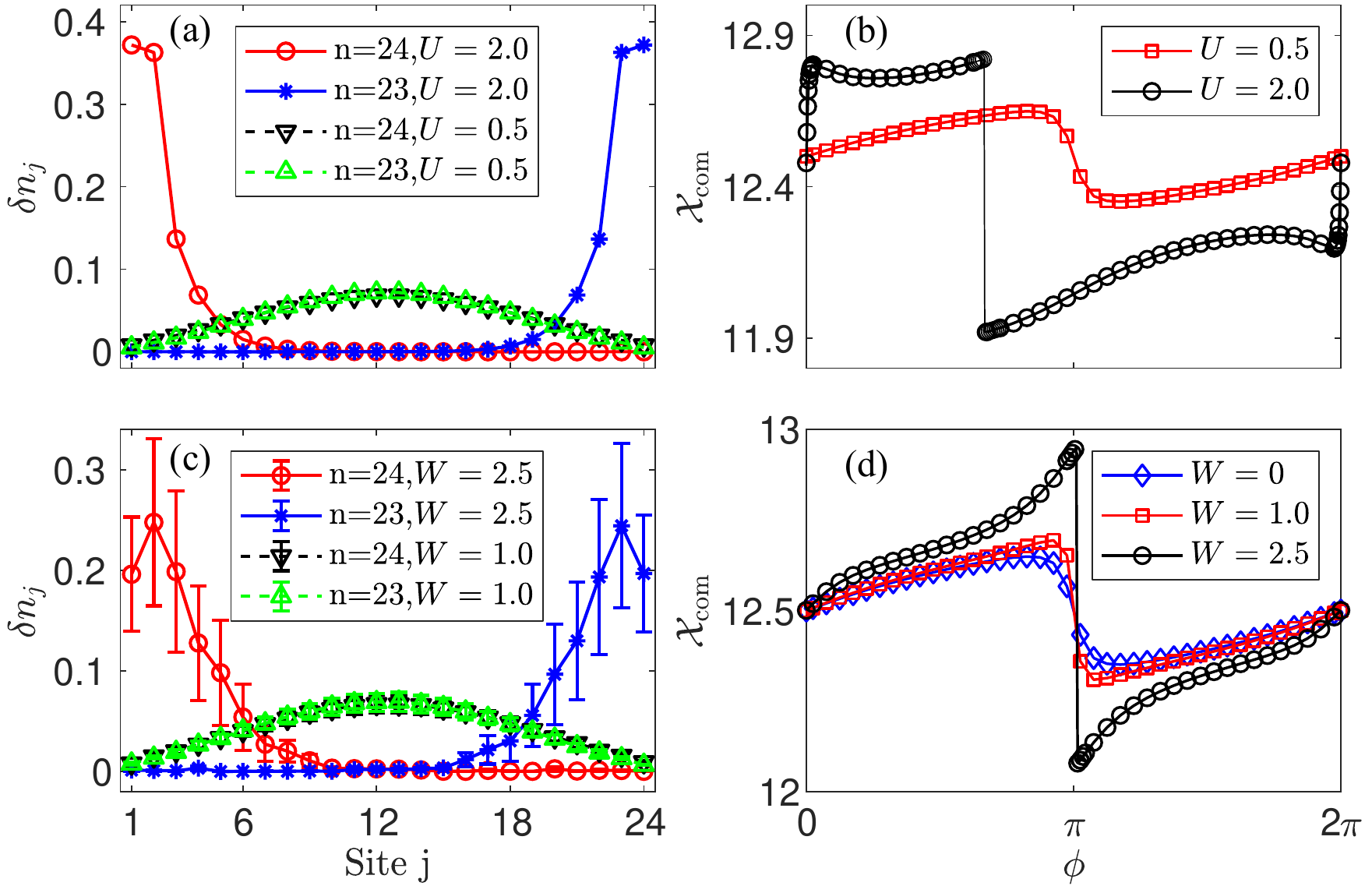}
\caption{(Color online) Quasiparticle density distribution $\delta n_j$ and center-of-mass $\mathcal{X}_{\mathrm{com}}(\phi)$ for $W=0$ (a,b) and $U=0.5$ (c,d) for $L=24$ from the DMRG and parameters in Fig. \ref{fig1}.}
\label{fig4}
\end{figure}

{\color{blue}\textit{DCTI and topological phase transitions.---}}For finite interaction and disorder, we compute $\gamma$ as a function of $W$ and $U$ in Fig.~\ref{fig3}(a) for $m_z=3$ and $L=8$ by the ED, which indicates three regimes for topologically trivial and nontrivial phases. Note that $\gamma$ is not well quantized in the region of $W\gtrsim3$ and $0\leqslant U\lesssim1$ due to the finite size effect in disorder configurations. To reduce the uncertainty of the phase boundary in this region, we recompute $\gamma$ by the DMRG for $L=24$ and the ED for $L=610$ when $U=0$, with the results plotted in Fig. \ref{fig1}(b). Moreover, we justify the phase boundary by the unsupervised machine learning and anomaly detection of the entanglement spectrum \cite{PhysRevX.6.041033} and the local spin density \cite{Note1}.

We also confirm our results by gap and entanglement calculations. The half-chain entanglement $\mathcal{S}$ is defined as the von Neumann entropy \cite{PhysRevX.6.041033}: $\mathcal{S}=-\mathrm{Tr}_A[\rho_A\ln\rho_A]$ with $\rho_{A}=\mathrm{Tr}_B|\Psi^g\rangle\langle\Psi^g|$ being the reduced density matrix of two halves labeled by $A$ and $B$. As shown in Fig.~\ref{fig3}(b), the large $S$ indicates the topological phase that is highly entangled and consists with the result in Fig.~\ref{fig3}(a). We also compute the fidelity $\mathcal{F}$ of the ground state wave function against disorder, which is defined as
\begin{equation}\label{fidelity}
\mathcal{F}=\frac{2}{N_d(N_d-1)}\sum_{i\neq j}\braket{\Psi^g(\varphi_i)|\Psi^g(\varphi_j)},
\end{equation}
with $\varphi_{i,j}$ the random phases chosen from $N_d$ disorder samples. As shown in Fig.~\ref{fig3}(c), the fidelity $\mathcal{F}$ keeps nearly $1$ in the ordered phase, which is well separated from the disordered phase with $\mathcal{F}<1$.

Now we can find three phases under disorder and interaction: the ordered trivial BI for small $U$ and $W$ connecting to the $U=W=0$ limit, the DCTI connecting the TAI and TMI as a generalization of single-particle TAIs to many-body interacting regime, and the disordered trivial phase for large $W$. There are three typical topological phase transitions with increasing $U$ or $W$, as shown in Figs.~\ref{fig3}(d-f). For the first case with increasing $U$ and fixed $W=2$ in Fig.~\ref{fig3}(d), the topological invariant changes from $\gamma=0$ to 1 with the excitation gap closing ($\Delta_g\approx0$) at the critical interaction strength. Near the critical point, $\mathcal{F}$ decreases sharply from unity and $\mathcal{S}$ increases quickly. These results signal the topological phase transition from an ordered trivial BI to the DCTI. For the second case with increasing $W$ and fixed $U=0.5$ in Fig.~\ref{fig3}(e), the disorder-induced topological phase transition from the BI to the DCTI is also indicated by the change of $\gamma$ and other quantities. However, with further increasing $W$, the DCTI becomes a disordered trivial phase without gap closing. For the third case with increasing $W$ and fixed $U=2$ in Fig.~\ref{fig3}(f), the transition from the DCTI to the trivial phase is well indicated by the sharp change of the values of $\gamma$, $\mathcal{S}$ and $\mathcal{F}$, but without gap closing. By repeating the procedure of determining the transition points at different values of $U$ and $W$ from the ED and DMRG, we map out the phase diagram in Fig. \ref{fig1}(b). We find that the first phase boundary is accompanied by the discontinuity of the magnetization \cite{Note1}.

{\color{blue}\textit{Unified analysis of phase boundaries.---}}Based on the MF and SCBA mehtods, we develop a unified approach to obtain the phase boundaries between the trivial BI and different topological phases. Under the Hartree-Fock MF approximation, the density-density interaction term in Hamiltonian (\ref{Ham}) can be linearized as $n_{j\sigma}n_{j+1\sigma}\approx\braket{n_{j\sigma}}n_{j+1\sigma}+\braket{n_{j+1\sigma}}n_{j\sigma}-\braket{n_{j\sigma}}\braket{n_{j+1\sigma}}$.
We find that the interaction term in the BI phase can be further simplify to $\sum_{j\sigma}n_{j\sigma}n_{j+1\sigma}\approx\rho_sU\sum_{j}(n_{j\uparrow}-n_{j\downarrow})$ up to an irrelevant constant with $\rho_s$ being the spin density difference (the spin density distribution of the ground state is nearly site-independent) \cite{Note1}. Thus, the interaction effectively renormalizes the Zeeman term. By numerically determining $\rho_s$, we obtain the interaction-renormalized Zeeman strength $\tilde{m}_z\approx m_z-U/1.14$ \cite{Note1}. In the clean limit, the interaction-induced TMI emerges when $|\tilde{m}_z|<2$, which agrees well with the topological phase boundary shown in Fig.~\ref{fig2}(a). The analysis works in the ordered BI regime where the disorder is not dominated. The disorder effect can be accounted based on an effective medium theory and the SCBA method \cite{Groth2009,HJiang2009,CZChen2015}.
The key of the SCBA method is to self-consistently obtain the self-energy introduced by the disorder, and then to include the self-energy as renormalization to the clean Hamiltonian. For our model, the self-energy term $\Sigma(W)$ satisfies the self-consistent equation \cite{Note1}
\begin{equation}
\frac{1}{E_f-H_{\text{MF}}(k)-\Sigma(W)}=\langle\frac{1}{E_f-H_{\text{eff}}(k,W_q)}\rangle_q,
\end{equation}
where $E_f\equiv0$ is the Fermi energy, $H_{\text{MF}}(k)=(\tilde{m}_z-2t\cos k)\sigma_z-2t_{s}\sin k\sigma_y$ with $\sigma_{y,z}$ the Pauli matrices is the clean MF Hamiltonian, $H_{\text{eff}}$ denotes the effective Hamiltonian renormalized by the disorder $W_q=W\cos(2\pi\alpha q)$ with index $q=1,2,...,N_q$, and $\braket{\cdots}_q$ denotes averaging over $N_q$ disorder samples. Here the self-energy is given by $\Sigma=\Sigma_z\sigma_z+\Sigma_y \sigma_y$. After numerically obtaining $\Sigma(W)$ for given $W$ \cite{Note1}, the Zeeman strength $\tilde{m}_z$ and the hopping strength $t_s$ are renormalized according to $\bar{m}_z=\tilde{m}_z+\Sigma_z\approx m_z-U/1.14+\Sigma_z$ and $\bar{t}_s=t_s+\Sigma_y$. This produces the topological phase boundaries at $|\bar{m}_z|=2$, which are plotted as the red dashed lines in Figs.~\ref{fig1}-\ref{fig3} and agree well with the numerical results. Thus, the topological phase transitions from the trivial BI to topological phases can be unifiedly attributed to the interaction and disorder renormalization on the Zeeman strength.

{\color{blue}\textit{Proposed detections.---}}Finally, we propose that the correlated topological phases may be detected from the disorder/interaction induced edge excitations and charge pumping \cite{Thouless1982} in the OL. %The appearance of in-gap boundary states is a hallmark of topological phases.
The density distribution of the quasiparticle excitation when a fermion is added to the lattice filled by $n$ fermions can be defined as $\delta n_j=\braket{\psi^g_{n+1}|n_j|\psi^g_{n+1}}-\braket{\psi^g_{n}|n_j|\psi^g_{n}}$~\cite{HGuo2011},
where $\ket{\psi^g_{n}}$ is the ground state of filling $n$ fermions under OBC, and $n_j=n_{j\uparrow}+n_{j\downarrow}$. We compute $\delta n_j$ for $L=24$ and $n=23,24$ by the DMRG in Figs. \ref{fig4}(a,c). The interaction and disorder can induce two edge modes when the system is driven from the trivial BI to the TMI and DCTI, respectively. So the filling $N_a=23,24,25$ is excepted for two edge collective excitations, which are different from zero-energy single-particle modes in the $U=0$ case (related to $g$ \cite{Note1}). The edge excitations can be probed by site-resolved spectra of atomic density distributions \cite{Leseleuc2019}.

The topological pumping allows a quantized particle transport when adiabatically encircling in parameter space~\cite{Thouless1982,Atala2013}. We design a pump scheme by tuning the parameters in Hamiltonian~(\ref{Ham}):
$\pm t_{s}\rightarrow\pm t_{s}(1+\sin\phi)$ and $m_z\rightarrow m_z+0.5\cos\phi$ via the phase $\phi$ varying from $0$ to $2\pi$. The charge pumping can be observed from the center-of-mass of an atomic gas with respect to the adiabatic parameter~\cite{Lohse2015,Nakajima2016,HILu2016,Schweizer2016,Kuno2020,Nakajima2021}: $\mathcal{X}_{\mathrm{com}}(\phi)=\frac{1}{N_a}\sum_{j}j\braket{\Psi^g(\phi)|n_j|\Psi^g(\phi)}$ under OBC.
As shown in Figs.~\ref{fig4}(b) and \ref{fig4}(d), when the system is driven from the BI to the TMI or DCTI, the center of mass $\mathcal{X}_{\mathrm{com}}(\phi)$ shows a jump. The obtained pumped charge $Q=\int_0^{2\pi}d\phi\frac{\partial\mathcal{X}_{\mathrm{com}}}{\partial\phi}=0$ for the trivial cases, and $Q\approx1$ for the topological phases in the large $L$ limit \cite{Note1}. The atomic pump in OLs has been observed from center-of-mass measurements \cite{Lohse2015,Nakajima2016,HILu2016}, even in the presence of interactions \cite{Schweizer2016} or disorders \cite{Nakajima2021}. %Our results indicate that the bulk-boundary correspondence, which involves the topological invariants/charge pumping and edge states, holds for the disorder/interaction induced topological phases.

{\color{blue}\textit{Conclusion.---}}In summary, we have demonstrated that the TAI and TMI phases can emerge and be connected without gap closing in a 1D disordered OL of interacting fermions. We have uncovered the DCTI phase induced by the combination of disorder and interaction. We have developed a unified theory to obtain the phase boundaries between the trivial BI and different topological phases. The predicted topological phases could be detected from edge excitations and charge pumping in OLs. Our work reveals the first many-body generalized TAI and provides a unified picture of the disorder and interaction driven topological phase transitions. These results show a unified framework to explore the seemingly different TAIs and TMIs with the intermediate DCTI phase, which is applicable to other models and higher dimensional systems. It would be also interesting to further study the interplay of many-body localization and topology.

\textit{Note added:} Recently, we noticed a related work on the symmetry-protected topological phase of hard-core bosons in 1D Rydberg atom chains with structural disorders \cite{arXiv2104}.

\acknowledgments
This work was supported by the National Natural Science
Foundation of China (Grants No. U1830111, No. 12047522, No. 12074180, and No. U1801661), the Key-Area Research and Development Program of Guangdong Province (Grant No. 2019B030330001), the Science and Technology of Guangzhou (Grants No. 2019050001), and the Guangdong Basic and Applied Basic Research Foundation (Grants No. 2020A1515110290 and No. 2021A1515010315).

\bibliography{reference}

\clearpage

\onecolumngrid
\appendix

\section{Supplemental Materials for \\Connecting Topological Anderson and Mott Insulators in Disordered Interacting Fermionic Systems}

\subsection{Finite-size scaling}
In this part, we present finite-size scalings of the topological invariant, excitation gap, charge pumping, and give the numerical evidence of the ground state degeneracy in the topological regions. We first show the Berry phase $\gamma$ calculated in our DMRG simulation up to $L=48$ sites, and establish this quantized topological invariant can preserve in the large-$L$ limit by extrapolation.

\begin{figure}[!h]
\centering
\includegraphics[width=0.95\textwidth]{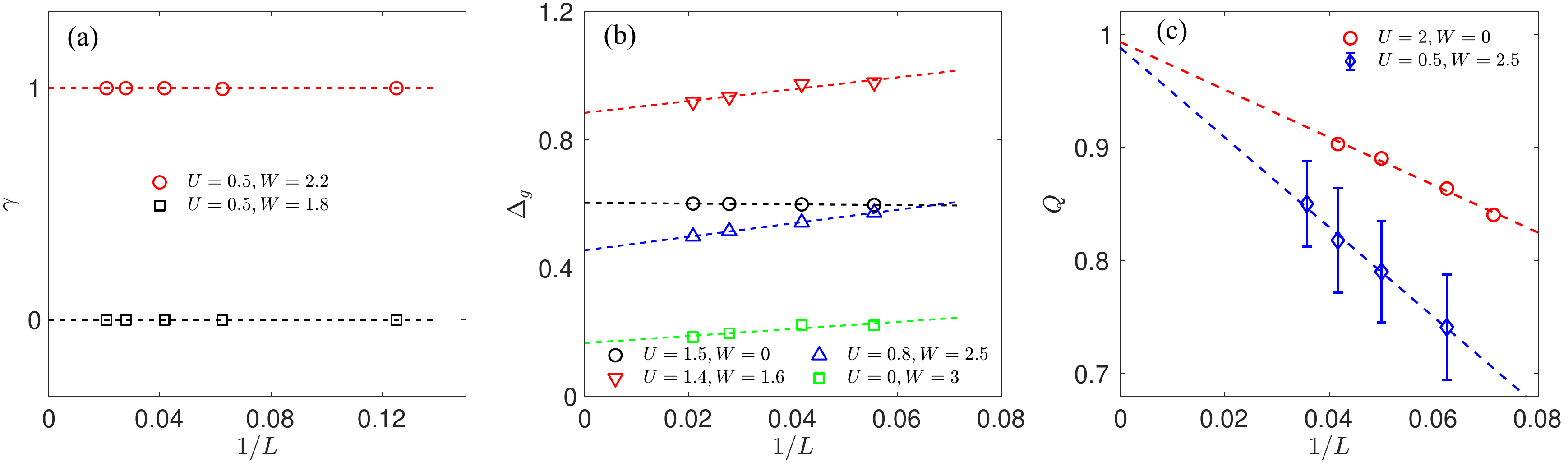}
\caption{(Color online) Finite-size scalings of (a) the Berry phase $\gamma$, (b) excitation gap $\Delta_g$, and (c) total pumped charge $Q$. The parameters are $t=1$, $t_{s}=0.95$, $m_z=3$, $\alpha=(\sqrt{5}-1)/2$, and offset phase $\varphi=0$ for (a) and $20$ offset phases are averaged for $W\neq 0$ in (b) and (c). Dashed lines are linear fits of the numerical data obtained from the DMRG.}
\label{figS1}
\end{figure}

In Fig.~\ref{figS1}(a), the finite-size scaling of the Berry phase $\gamma$ is plotted as a function of $1/L$ in the topological ($W/t=2.2$) and trivial ($W/t=1.8$) regions near the phase transition boundary. The twist phase interval $[0,2\pi]$ is equally cut into 24 parts, the offset phase is fixed to $\varphi=0$. We have checked the convergence of the Berry phase with the twist phase interval cut into $40$ parts. In each DMRG simulation, we keep up to 600 Schmidt values and use a convergence criterion $|\Delta E_g|/E_g<10^{-8}$ of the ground state (and the first excited state if computed). These criteria are also used in other DMRG simulations involved in this work. We further calculate the excitation gap in topological phases (TMI, DCTI, and TAI) under PBC and do the finite-size scaling to see whether the gap can preserve in the thermodynamic limit. Figure~\ref{figS1}(b) shows the excitation gap scalings of four typical parameters in the topological region, and the linear fits indicate that $\Delta_g$ has finite value when $L\rightarrow\infty$.

We present finite size scalings of the total pumped charge $Q$ obtained from the DMRG simulations, as shown in Fig.~\ref{figS1}(c). In the TMI phase ($U=2,W=0$), the pumped charge $Q$ for $L\rightarrow\infty$ tends to $Q\approx0.993$ from the linear fit of the numerical data. For the DCTI phase ($U=0.5,W=2.5$), we numerically obtain $Q\approx 0.988$ from the scaling when $L\rightarrow\infty$.

\subsection{Ground state degeneracy related to edge modes near half filling}

\begin{figure}[!h]
\centering
\includegraphics[width=0.9\textwidth]{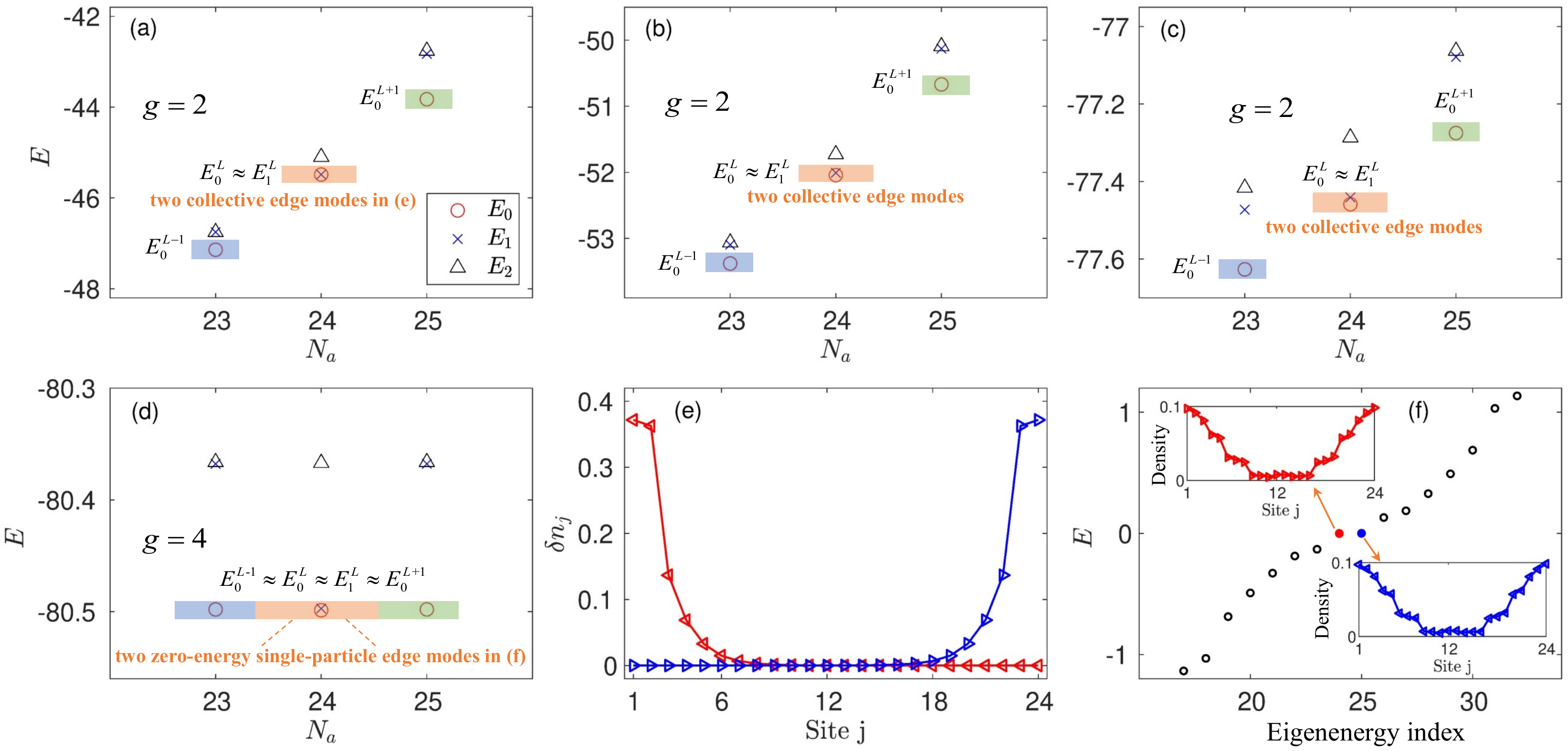}
\caption{(Color online) Energies $E$ of lowest three states near the half filling under OBC with total atomic numbers $N_a=L-1,L,L+1$ for (a) the TMI with $U=2$ and $W=0$, (b) the DCTI with $U=1.5$ and $W=0.5$, (c) the DCTI with $U=0.2$ and $W=3$, and (d) the TAI with $U=0$ and $W=3$, Parameters are $t=1$, $t_{s}=0.95$, $m_z=3$, $\alpha=(\sqrt{5}-1)/2$, $L=24$, and 20 different offset phases are averaged when $W\neq0$. (e) The density distribution of two collective excitation edge modes (see Fig. 4(a) and the discussion in the main text). (f) The single-particle spectrum and the density distribution of two zero-energy single-particle edge modes. The results in (a-e) are obtained from the DMRG, and that in (f) is obtained from the ED under the single-particle representation.}
\label{figS2}
\end{figure}

We compute the three lowest eigenenergies $E_{0,1,2}^{N_a}$ of the many-body system near half filling under OBC from the DMRG for the particle number $N_a=L-1,L,L+1$ with fixed $L=24$. The numerical results for four typical cases are shown in Fig. \ref{figS2}(a-d). The ground state degeneracy $g$ is refereed to counting the number of degenerate energies among four energies $\{E_0^{L-1},E_0^{L},E_1^{L},E_0^{L+1}\}$. In the trivial phase, we obtain the single many-body ground state without degeneracy ($g=1$ and not shown here). For the topological phases of interacting fermions [the TMI and the DCTI with $U\neq0$], one can find $E_0^{L-1}<E_0^{L}\approx E_1^{L}<E_0^{L+1}$ from Figs. \ref{figS2}(a-c), such that the ground state degeneracy in these cases is $g=2$. Here the twofold degeneracy from $E_0^{L}\approx E_1^{L}$ under OBC corresponds to two topological collective excitations for the filling $N_a=L-1,L,L+1$, which are occupied on each edge in this finite system of $L=24$, as shown in Fig. \ref{figS2}(e) and Fig. 4 in the main text. For the noninteracting topological phase of fermions (the TAI with $U=0$ and $L=24$) in Fig. \ref{figS2}(d), one has $E_0^{L-1}\approx E_0^{L}\approx E_1^{L}\approx E_0^{L+1}$ and thus obtain the degeneracy $g=4$. Here $E_0^{L}\approx E_1^{L}$ also corresponds to two edge modes as in the interacting case, and the energy difference between $E_{0,1}^{L}$ and $E_{0}^{L\pm1}$ gradually decreases and tends to zero as the interaction strength $U\rightarrow0$. In this noninteracting limit, one thus has fourfold ground state degeneracy, which indicates that the two edge modes are zero energy and can be either empty or occupied. The two zero-energy edge modes within the single-particle spectrum are numerically confirmed by the ED under the single-particle representation, as shown in Fig. \ref{figS2}(f). Another difference between single-particle and collective excitation edge modes is their density distributions, which can been clearly seen from Figs. \ref{figS2}(e,f). In contrast to two collective excitations occupied on each edge in the interacting topological phases,  for the noninteracting topological case, one fermion can occupy (and transfer between) two edges owing to the hybridization for finite chains. The differences of edge modes in the noninteracting and interacting (symmetry-protected) topological phases closely related to ground state degeneracy under OBC have been noted in Ref. \cite{XZhou2017} and moreover experimentally observed in Ref. \cite{Leseleuc2019}.

\subsection{Machine learning the phase diagram in a unsupervised and automated fashion}

Recently, machine learning methods are emerging as a versatile toolbox to explore the quantum phase diagrams~\cite{vanNieuwenburg2017,Rodriguez-Nieva2019,PhysRevLett.125.170603,Kaeming2021}. Especially, the unsupervised learning with anomaly detection has showcased its exceptional applicability in both theory~\cite{PhysRevLett.125.170603} and experiments~\cite{Kaeming2021}. In this part, we apply the unsupervised learning and anomaly detection proposed in Ref.~\cite{PhysRevLett.125.170603} to benchmark the ground-state phase diagram of our model.

%In this part, we explore the phase diagram of quasiperiodic random disorder model via machine learning (ML). Recent ML methods are emerging as a versatile toolbox to explore exotic quantum phase diagrams~\cite{vanNieuwenburg2017,PhysRevLett.120.066401,PhysRevB.103.035413,PhysRevA.103.012419,Rodriguez-Nieva2019,PhysRevLett.124.185501,PhysRevLett.124.226401,Dawid_2020,PhysRevLett.125.170603,PhysRevLett.125.225701}. Among of them, unsupervised learning, especially anomaly detection has showcased its exceptional applicability in both theory~\cite{PhysRevLett.125.170603} and experiments~\cite{kaming2021unsupervised}. Here, we apply anomaly detection proposed in Ref.~\cite{PhysRevLett.125.170603} to benchmark our model.

\begin{figure}[!h]
\centering
\includegraphics[width=0.95\textwidth]{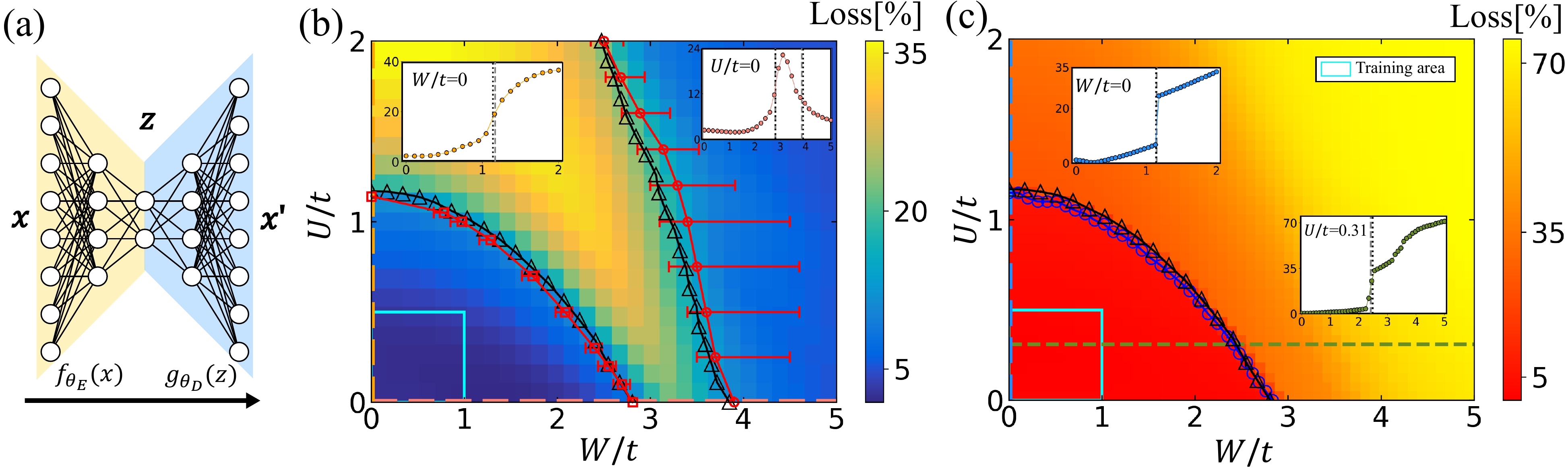}
\caption{(Color online) (a) Schematic of the neural network autoencoder with encoder and decoder. The input data $\boldsymbol x$ are mapped to the low bottleneck $\boldsymbol z$ via encoder $\boldsymbol z=f_{\theta_E}( \boldsymbol x)$ and decoded to $\boldsymbol x'=g_{\theta_D}(\boldsymbol z)$. The 2D loss map of the mean loss of the autoencoder after training in the parameter space $(U/t,W/t)\in[0,0.5]\times[0,1]$ with (b) the entanglement spectrum and (b) the local spin density as input data, respectively.
The insets show the loss along the dashed lines. Vertical gray dashed (black dotted) lines in the insets are the transitions obtained by the anomaly detection (the ED and DMRG simulations). Two different phases are distinguished via different plateaus of loss value. The black triangle and blue circle symbols in (b) and (c) guided by the solid line denotes the topological phase boundaries determined by using anomaly detection. Red circle and square symbols with error bars for the uncertainty guided by the red solid lines are the phase boundaries determined by the ED and DMRG, as plotted in Fig. 1(b) in the main text.}
\label{figS3}
\end{figure}

We apply the neural network autoencoder for anomaly detection. An autoencoder is a type of neural network that
consists of two parts as shown in Fig.~\ref{figS3}(a). The encoder part takes the high dimensional input data $\boldsymbol x$ and maps it to a low latent variable $\boldsymbol z$ via a parametrized function $\boldsymbol z=f_{\theta_E}(\boldsymbol x)$. The decoder part takes the latent variable $\boldsymbol z$ and maps it back to $\boldsymbol{x}'=g_{\theta_D}(\boldsymbol z)$. The goal of the autoencoder is to find an efficient representation of the input data $\boldsymbol{x}$ through the encoder at the bottleneck, from which the decoder can reproduce the input data $\boldsymbol{x'}$. The parameters $\theta_E$ and $\theta_D$ are trained via the minimization of the mean squared error loss function $\bar{L}(\boldsymbol x,\boldsymbol x')=||\boldsymbol x-\boldsymbol x'||^2$, which is averaged over 20 disorder configurations in our case.

The main idea of anomaly detection scheme is that after training the autoencoder in a region of the phase diagram, it learns the characteristic features of the phase where the data were taken from and fails to reproduce data
from the other phases. Therefore, the failure leads to a higher loss, from which we
deduce that the data do not belong to the same phase as the ones used to train the autoencoder. By looking at the loss for all data after training in only part of the phase diagram, we are able to distinguish different phases via different plateaus of the loss
function. Furthermore, by fitting the loss curve and finding the transition points, we obtain phase boundaries.
Below we use this method based on the autoencoder to benchmark the phase diagram of the model shown in Fig. 1(b).

In order to explore the topological properties of model, we first use the entanglement spectrum \cite{vanNieuwenburg2017} as our input data $\boldsymbol x$ to explore the phase diagram of our model. We note that half-chain entanglement has already indicated the regimes for the topologically trivial and the nontrivial phases, as shown in Fig.~2(b) in the main text. The entanglement spectrum also plays an important role in the characterization of many-body quantum systems and is experimentally accessible in cold atom systems~\cite{PhysRevX.6.041033}. We divide the whole system into two equal subsets $A$ and $B$, after which the reduced density matrix of subset $A$ is calculated by partially tracing out the degrees of freedom in $B$: $\rho_{A}=\mathrm{Tr}_B|\Psi^g\rangle\langle\Psi^g|$. Denoting the eigenvalues of $\rho_A$ as $\lambda_i$, the entanglement spectrum is defined as the set of numbers $-\ln{\lambda_i}$. The input data $\boldsymbol x$ are $32\times8$ matrices, each row of which consist of the lowest 8 eigenvalues of $-\ln{\lambda_i}$ for 32 different twist boundary phases $\theta$s. Here the autoencoder is a convolutional neural network and the encoder part consists of two convolution layers, which have 16 kernels of size $2\times2$ with 2 strides, 8 kernels of size $2\times2$ with 2 strides, respectively. The input data are then transformed into $\boldsymbol z$ as $8\times2$ matrices with 8 channels. Following the decoder part are two upsampling convolutional layers both have 16 kernels of size $2\times2$ with 2 strides. Finally, a convolutional layer with 1 kernel of size $2\times2$ is used to reproduce the output with the same dimension as the input data $\boldsymbol x'$.

Assuming no a prior knowledge, we start by training data points at the parameter space $(U/t,W/t)\in[0,0.5]\times[0,1]$ with system size $L=8$, which is the ordered trivial BI phase according to Fig.~2(a) in the main text. By testing with data points from the whole phase diagram, we can clearly distinguish the boundaries between the trivial phase and the non-trivial phase in Fig.~\ref{figS3}(b). The different loss levels and the sharp transitions at the phase boundaries indicate two different phases.

We also use another experimentally accessible quantity, i.e., the local spin density $\braket{n_{j\sigma}}$ to reveal the phase transitions from the trivial BI to the topological phases. The input data of the local spin density are $2L\times1$ matrices of the form $[\braket{n_{1\uparrow}},\braket{n_{1,\downarrow}},\braket{n_{2\uparrow}},\braket{n_{2\downarrow}},...,\braket{n_{L\uparrow}},\braket{n_{L\downarrow}}]$. We numerically generate the data for each pair of parameters over 20 quasiperiodic configurations with respect to the offset phase $\varphi$. Now the autoencoder is taken as a fully connected neural network consisting of three hidden layers which have 8, 4 and 8 neurons, respectively. We take the training data points at the parameter space $(U/t,W/t)\in[0,0.5]\times[0,1]$, which is the ordered trivial BI. After training, we can obtain the boundaries between the trivial BI to the topological phases.

Overall, the phase boundaries predicted by this unsupervised and automated machine learning match well with those obtained from the ED and DMRG results of the phase diagram in Fig. 1(b) in the main text. The right side phase boundary from nontrivial to trivial insulator determined by the ED and DMRG methods suffer significant uncertainty due to the disorder fluctuations and the finite-size effect, especially when $0\leqslant U\lesssim1$ and $W\gtrsim3$ (here the DMRG for $L=N_a=24$ and $U\neq0$ and the ED for $L=610$ for $U=0$ has been used to reduce the uncertainty). Surprisingly, the anomaly detection based the neural network method can indicate the phase boundary well in this uncertain region even thought the network is trained by the data obtained from $L=N_a=8$ small systems.

\subsection{Mean-field and self-consistent Born approaches}
\begin{figure}[!h]
\centering
\includegraphics[width=0.75\textwidth]{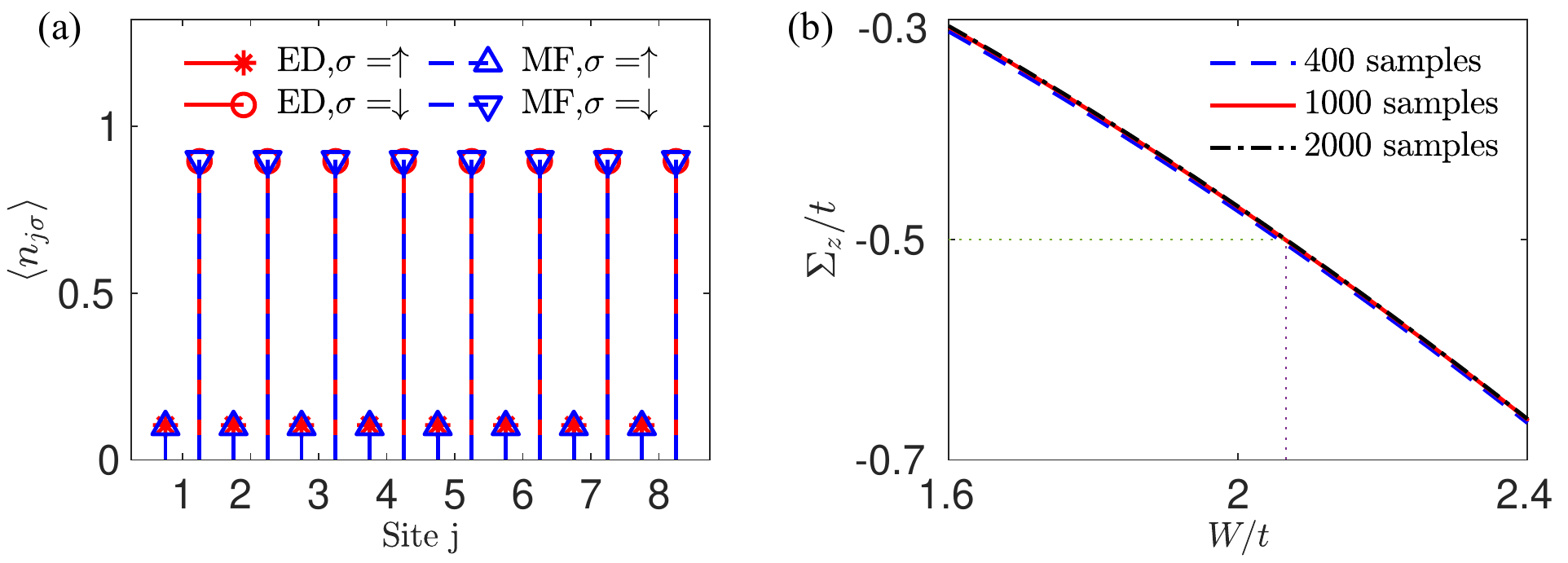}
\caption{(Color online) (a) Density distribution of each spin component for ED and MF approximated ground states in the clean limit. Parameters are chosen as $t=1$, $t_{s}=0.95$, $m_z=3$, $U=1$, and $L=8$. (b) The self-energy parameter $\Sigma_z$ as a function of $W$ for results averaged from $400,1000,2000$ number of disorder samples. Dotted lines indicate the critical disorder $W_c$ for $\tilde{m}_z+\Sigma_z=2t$. Parameters are $t=1$, $t_{s}=0.95$, $m_z=3$, $U=0.57$, $\alpha=(\sqrt{5}-1)/2$, and $\varphi=0$.}
\label{figS4}
\end{figure}

In this part, we describe some details of the unified analysis of the topological phase transitions from the trivial BI driven by the disorder and interaction. The density-density interaction term can be linearized by the Hartree-Fock MF approximation~\cite{YXWang2019}
\begin{equation}\label{eqmf}
{n}_{j\sigma}{n}_{j+1\sigma}\approx\braket{n_{j\sigma}}{n}_{j+1\sigma}+\braket{n_{j+1\sigma}}{n}_{j\sigma}-\braket{n_{j\sigma}}\braket{n_{j+1\sigma}}.\end{equation}
In the trivial BI phase, the fermion density of each spin component of the ground state is approximately site-independent, and the Hartree-Fork MF approximation can be simplified to
\begin{equation}
U\sum_{j=1}^L{n}_{j\sigma}{n}_{j+1\sigma}\approx\frac{2N_\sigma U}{L}\sum_{j=1}^L{n}_{j\sigma}+\mathrm{const.},\end{equation}
where $N_\sigma$ is the total occupation of spin-$\sigma$ fermions. We plot the density distributions of fermions calculated from both ED method and Hartree-Fork MF approach in Fig.~\ref{figS4}(a), which shows that the spin density is indeed site-independent and the MF ground state fairly captures the exact ground state property even near the trivial BI phase boundary ($U/t=1$). In this case, the density difference between spins $\braket{n_{\uparrow}}<\braket{n_{\downarrow}}$. By defining the density difference $\rho_s\equiv(N_\downarrow-N_\uparrow)/L>0$, the interaction term under the MF approximation can be further expressed as
\begin{equation}
\sum_{\sigma=\{\uparrow,\downarrow\}}\frac{2N_\sigma U}{L}\sum_{j=1}^L{n}_{j\sigma}=\sum_{j=1}^L[-\rho_s U(n_{j\uparrow}-n_{j\downarrow})+U(n_{j\uparrow}+n_{j\downarrow})]=UN_a-\rho_s U \sum_{j=1}^L(n_{j\uparrow}-n_{j\downarrow}),
\end{equation}
where $\rho_s$ affects the $\sigma_z$ component of the model and the interaction $U$ can renormalize the Zeeman strength as~\cite{YXWang2019}
\begin{equation}\label{eqrmz}
\tilde{m}_z=m_z-\rho_s U,
\end{equation}
up to an irrelevant constant energy $UN_a$. As is shown in Fig. \ref{figS4}(a), we numerically determine $\rho_s\approx 1/1.14$. We apply Eq.~(\ref{eqrmz}) to accurately characterize the phase boundary of the trivial BI and TMI in the clean limit.

As is discussed in the main text, the fidelity $\mathcal{F}$ of the ground state against disorder reveals that the trivial BI phase is ordered, and the above MF approach can be fairly applied. We further consider the MF relation in Eq. (\ref{eqrmz}) with nonzero disorder $W$ and include the neighboring density-density interaction $U$ into the Zeeman strength $m_z$ to analysis the topological phase boundary. The effects of disorder can be accounted as the self-energy from the effective medium theory and SCBA under as~\cite{Groth2009,HJiang2009,CZChen2015}
\begin{equation}\begin{split}
\frac{1}{E_f-H_{\mathrm{MF}}(k)-\Sigma(W)}=&\langle\frac{1}{E_f-H_\mathrm{eff}(k,W_q)}\rangle_q,\\
H_\mathrm{MF}(k)=&[\tilde{m}_z-2t\cos(k)]\sigma_z-2t_{s}\sin(k)\sigma_y,\\
H_\mathrm{eff}(k,W_q)=&[\tilde{m}_z-2t\cos(k)+W_q]\sigma_z-2t_{s}\sin(k)\sigma_y,
\end{split}\end{equation}
where $E_f$ is the Fermi energy and is set to zero in our model, $\sigma_{y,z}$ are Pauli matrices, $H_\mathrm{MF}(k)$ is the clean Hamiltonian under MF approximation, $\Sigma(W)$ is the self energy under the specify disorder strength $W$, $H_\mathrm{eff}(k,W_q)$ is the effective two-band disordered Hamiltonian with random parameter $W_q=W\cos(2\pi\alpha q)$ denoted by the index $q=1,2,...N_q$, and $\braket{\cdots}_q$ denotes averaging over $N_q$ disorder configurations.

Our numerical result indicates the self-energy has form $\Sigma=\Sigma_z\sigma_z+\Sigma_y \sigma_y$, with the $z$ component effectively renormalizes the Zeeman term $\bar{m}_z\rightarrow\tilde{m}_z+\Sigma_z$, and the $y$ component renormalizes the hopping paramter $\bar{t}_{s}\rightarrow t_{s}+\Sigma_y$ which is irrelevant to the topological phase transition. In Fig.~\ref{figS4}(b), we extract and plot $\Sigma_z$ as a function of the disorder strength $W$, which are averaged over different momenta $k$ and different sample numbers $N_q=400,1000,2000$. We find that $\Sigma_z$ tends to small value with increasing $W$, and the critical value $W_c$ is determined by the renormalized topological phase boundary $\tilde{m}_z+\Sigma_z=2t$ (e.g., $\tilde{m}_z=2.5$ for $U=0.57$). In practice, the self-energy converges quickly when increasing $N_q$ and we use $N_q=1000$ disorder samples in the main text.

\subsection{Discontinuous change of total magnetization at the topological phase transition}

\begin{figure}[!h]
\centering
\includegraphics[width=0.95\textwidth]{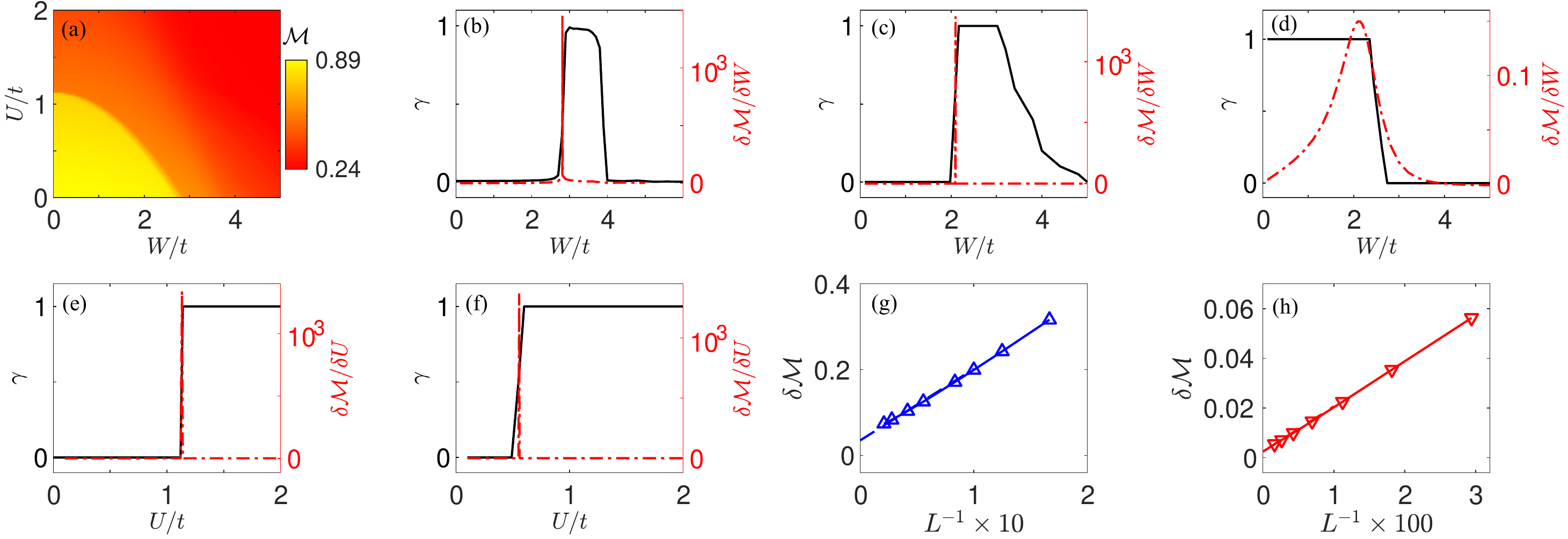}
\caption{(Color online) (a) The total magnetization $\mathcal{M}$ as a function of $U$ and $W$ for $L=N_a=8$ systems. A minus sign is added to $\mathcal{M}$ because $\mathcal{M}<0$ in our model. (b-f) The first derivative of $\mathcal{M}$ along with the Berry phase plotted as functions of $U$ or $W$ for $U=0$ (b), $U=0.5$ (c), $U=2.0$ (d), $W=0$ (e), $W=2$ (f); $L=610$ for $U=0$ and $L=8$ for $U\neq0$. (g,h) The finite-size scaling of the magnetization change $\delta \mathcal{M}$ near the phase transition point as functions of inverse system size $1/L$ for interaction-induced (g) and disorder-induced (h) phase transitions at $W=0$ and $U=0$, respectively. Other parameters are $t=1$, $t_{s}=0.95$, $m_z=3$, and $\alpha=(\sqrt{5}-1)/2$. (a,b,h) are obtained by the ED, and others are computed by the DMRG.}
\label{figS5}
\end{figure}

Here we numerically study the total magnetization $\mathcal{M}$ as a local order parameter to provide the first-order character of the quantum phase transition from the trivial BI to the topological phases. The total magnetization $\mathcal{M}$ in this spin-1/2 system is defined as
\begin{equation}
\mathcal{M}=\frac{1}{L}\sum_{j=1}^L\braket{\Psi^g|n_{j\uparrow}-n_{j\downarrow}|\Psi^g},
\end{equation}
where $\ket{\Psi^g}$ is the many-body ground state. In Fig.~\ref{figS5}(a), we plot $\mathcal{M}$ as a function of $W$ and $U$, which is obtained from the ED and averaged over $20$ random samples. The result shows a discontinuous change near the phase boundary (determined by other quantities as discussed in the main text) between trivial BI and the topological phases. To see the discontinuous change of $\mathcal{M}$ with respect to $W$ or $U$ more clearly, we further calculate the corresponding first derivatives $\delta \mathcal{M}(x)/\delta x$ with $x=U,W$. Numerically, the first derivative $\delta \mathcal{M}(x)/\delta x$ at $x=x_0$ up to the second order approximation is given by
\begin{equation}
\frac{\delta \mathcal{M}(x)}{\delta x}|_{x=x_0}\approx\frac{-\mathcal{M}(x_0+2\delta x)+8\mathcal{M}(x_0+\delta x)-8\mathcal{M}(x_0-\delta x)+\mathcal{M}(x_0-2\delta x)}{12 \delta x},
\end{equation}
and $\delta x=10^{-4}$ is chosen in the numerical calculation. In Figs.~\ref{figS5}(b-f), we plot the first derivative $\delta \mathcal{M}/\delta x$ as functions of $U$ or $W$ for five typical situations discussed in the main text. The interaction or disorder induced topological phase transitions from the trivial BI to the DCTI are accompanied by the divergence of $\delta \mathcal{M}/\delta x$ near the transition points. This indicates the disorder/interaction induced topological phase transitions have the character of first order phase transitions with discontinuous change of the total magnetization as a local order parameter. However, the transitions from the topological phases to the disordered trivial phase induced by stronger disorders shows no divergence of $\delta \mathcal{M}/\delta W$ [see Fig.~\ref{figS5}(d)].

To confirm that the discontinuity of the total magnetization can survive in the large $L$ limit, we perform the finite-size scaling of the magnetization change $\delta \mathcal{M}\equiv\mathcal{M}(x_c+\delta x)-\mathcal{M}(x_c-\delta x)$, which denotes the difference of the total magnet $\mathcal{M}$ between the right and left sides of the phase transition point $x_c$ (with the $\delta x=10^{-4}$). We present numerical results for the clean limit and the non-interacting limit in Figs.~\ref{figS5}(g) and \ref{figS5}(h), respectively. The extrapolations of $\delta \mathcal{M}$ to the $1/L\rightarrow 0$ limit indicate the discontinuous magnetization change can exist in the large $L$ limit. Thus, we argue that the first-order topological quantum phase transition induced by the disorder/interaction preserves in the thermodynamic limit.

\subsection{Results for the case of $m_z=4$ and the random disorder case}

\begin{figure}[!h]
\centering
\includegraphics[width=0.6\textwidth]{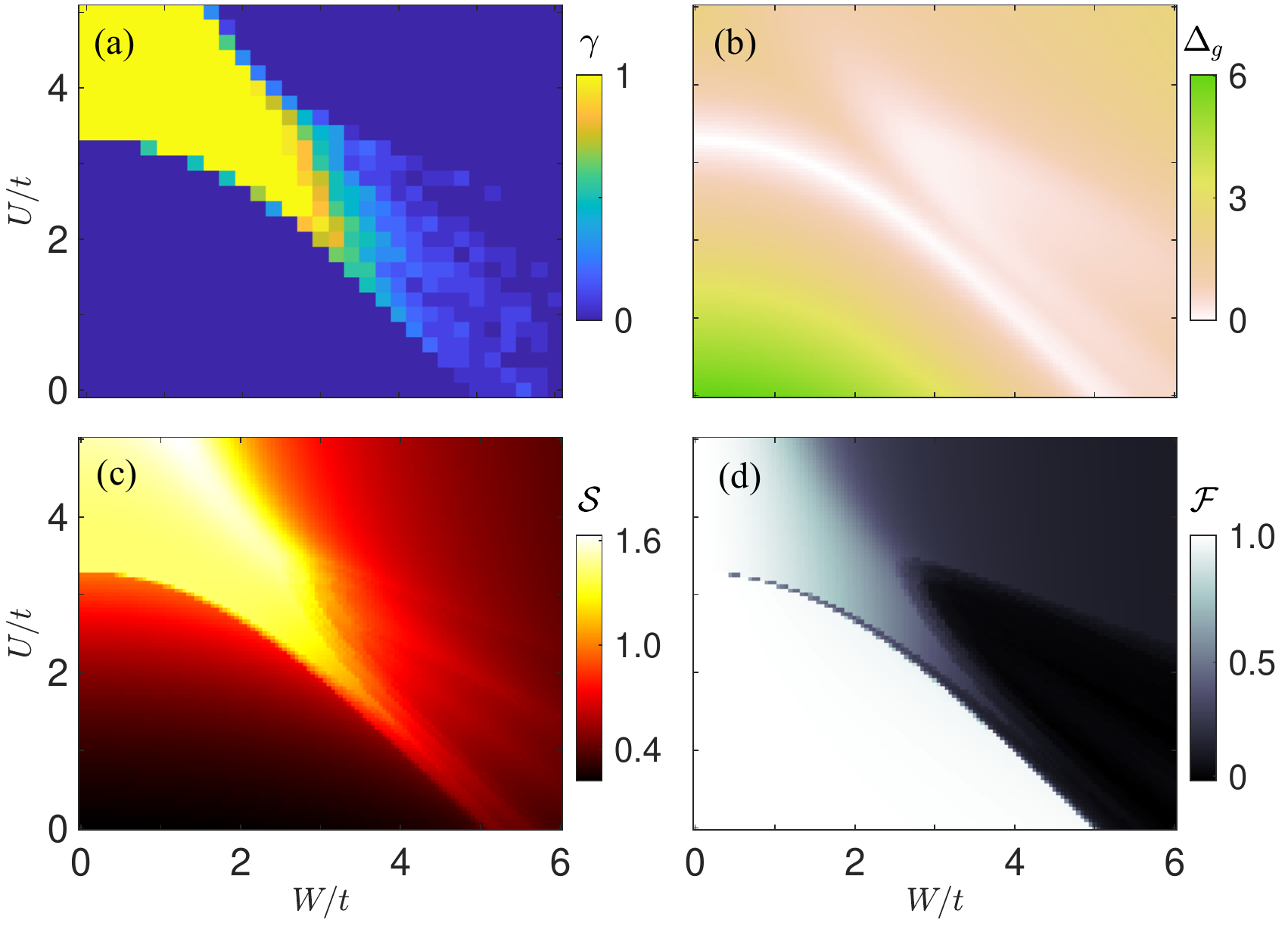}
\caption{(Color online) (a) Berry phase $\gamma$, (b) excitation gap $\Delta_g$, (c) half-chain entanglement $\mathcal{S}$, and (d)  ground-state fidelity $\mathcal{F}$ plotted as functions of $W$ or $U$ for $m_z=4$. All data are obtained from the ED for $L=8$ systems. Other parameters are the same to Fig. 1 in the main text.}
\label{figS6}
\end{figure}

We present numerical results of the quantized Berry phase $\gamma$ as functions of $W$ and $U$ for $m_z=4$ in Fig.~\ref{figS6}(a), in which case the DCTI phase induced by the combination of disorder and interaction remains while the noninteracting TAI phase is absent. The phase transitions from the trivial BI to the topological phases accompany with the gap closing as shown in Fig.~\ref{figS6}(b). The half-chain entanglement $\mathcal{S}$ and the ground-state fidelity $\mathcal{F}$ (against disorder) are plotted in Figs.~\ref{figS6}(c) and \ref{figS6}(d), respectively. The large entanglement signals the topological states are highly entangled and the fidelity suggests that trivial BI phase is ordered and the others are disordered. These numerical results consist with those for $m_z=3$, except the absence of disorder-induced TAIs in the noninteracting limit for $m_z=4$.

\begin{figure}[!h]
\centering
\includegraphics[width=0.9\textwidth]{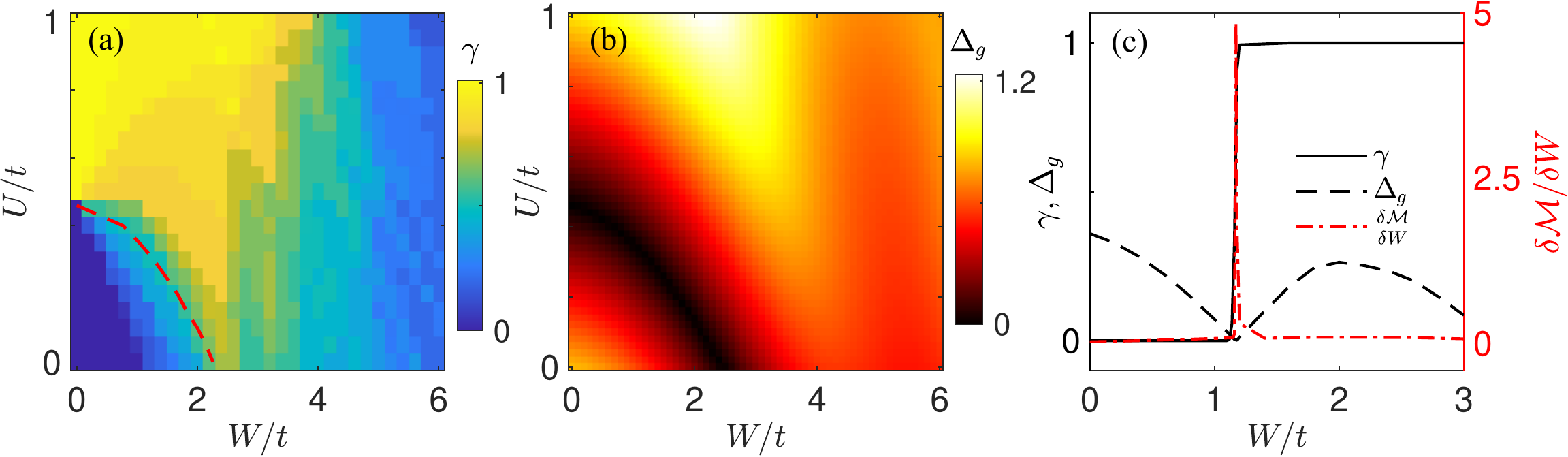}
\caption{(Color online) (a) Berry phase $\gamma$ and (b) excitation gap $\Delta_g$ on the $W$-$U$ plane for $m_z=2.4$ and $L=N_a=8$ obtained from the ED. The red dashed curve is the phase boundary obtained from the MF and SCBA approaches. The data in (a) are averaged over 20 random disorder realizations. (c) Berry phase $\gamma$, excitation gap $\Delta_g$, and first derivative of $\mathcal{M}$ as a function of $W$ for a disorder configuration with $U=0.25$, $m_z=2.4$ and $L=24$ obtained from the DMRG. Other parameters are $t=1$ and $t_{s}=0.95$.}
\label{figS7}
\end{figure}

We also consider the random disorder by replacing the quasiperiodic disorder $m_{j}=m_z+W\cos(2\pi\alpha j+\varphi)$ with the random disorder $m_{j}=m_z+W_j$ in the model Hamiltonian (1) in the main text, where $W_j$ is uniformly distributed in $[-W,W]$. We display the Berry phase $\gamma$ on the $W$-$U$ plane for $L=N_a=8$ and $m_z/t=2.4$ in the random disorder case, as shown in Fig.~\ref{figS7}(a). Although $\gamma$ is not well quantized near the phase boundary due to the finite size and fluctuation effects in this case, the disorder and interaction induced topological phase transitions from the trivial BI to the DCTI can also be observed. The red dashed curve is the phase boundary obtained from the MF and SCBA approaches. In Fig.~\ref{figS7}(b), the excitation gap $\Delta_g$ is displayed for a certain random disorder configuration, which clearly shows the gap closing near the topological phase transition from the BI to topological phases. To reduce the finite size effect of the random disorder, we further plot the DMRG result for a larger system with $L=N_a=24$ in Fig.~\ref{figS7}(c), where the Berry phase $\gamma$, excitation gap $\Delta_g$, and the first derivative of the total magnetization $\delta \mathcal{M}/\delta W$ as a function of $W$ are plotted. The Berry phase $\gamma$ is now well quantized in the DMRG simulation, and the results show the topological phase transition accompanied by the gap closing and the divergence of the first derivative of the total magnetization. These results are similar to those for the quasiperiodic disorder discussed in the main text. This indicates that the adiabatical connection between the TAI and TMI phases, the DCTI phase, and the unified analysis of the topological phase transitions in the main text preserve for both quasiperiodic and random disorders.

\subsection{Results for the case with the intra-leg interaction term}

\begin{figure}[!h]
	\centering
	\includegraphics[width=0.7\textwidth]{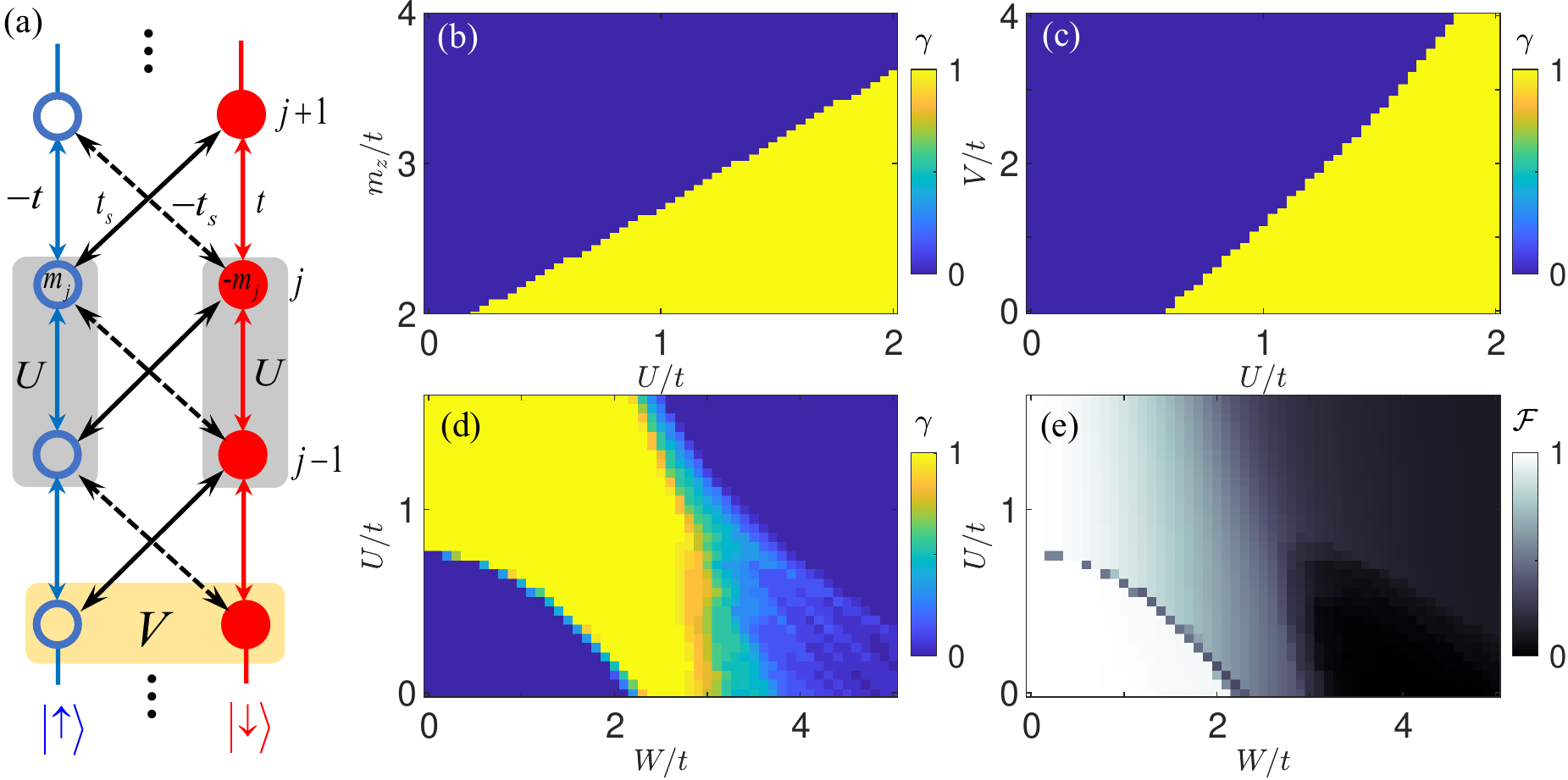}
	\caption{(Color online) (a) The proposed model with additional interaction $V$ between different spin species (legs) on the same site of the ladder. (b) Berry phase $\gamma$ on the $U$-$m_z$ plane for $W=0$ and $V=0.5$. (c) Berry phase $\gamma$ on the $U$-$V$ plane for $W=0$ and $m_z=2.5$. (d) Berry phase $\gamma$, and (e) fidelity $\mathcal{F}$ as functions of $W$ and $U$ for $V=0.5$ and $m_z=2.5$. All data are computed by the ED for $L=N_a=8$ and are averaged over 20 quasiperodic configurations of random offset phases. Other parameters are the same as those in Fig. 3(a) in the main text.}
	\label{figS8}
\end{figure}

We now show that our main results with the topological phase diagram can preserve in the presence of finite density-density interaction between different spin components along vertical bonds of the ladder, as shown in Fig.~\ref{figS8}(a). By adding this intra-leg (on-site) interaction term $V\sum_jn_{j\uparrow}n_{j\downarrow}$ in the Hamiltonian (1) in the main text, we numerically compute the Berry phase $\gamma$ and the fidelity $\mathcal{F}$ by the ED for several typical cases, with the results shown in Figs.~\ref{figS8}(b-e). In the absence of disorders with $W=0$ [Figs.~\ref{figS8}(b,c)], one can find the increasing of the interaction strength $U$ can induce the topological phase transition from the trivial phase ($\gamma=0$) to the TMI phase ($\gamma=1$) for finite $V$, although the increasing of $V$ will shrink the region of the TMI phase. In Figs.~\ref{figS8}(d,e), we plot $\gamma$ and $\mathcal{F}$ on the $W$-$U$ plane for $V=0.5$. We can find that under this addition interaction term, the disorder-induced TAIs for $U=0$ and DCTIs for $0< U\lesssim0.8$ from the trivial BI phase exhibit in the topological phase diagram, similar to the results shown in Fig. 3 in the main text.

\begin{figure}[!h]
	\centering
	\includegraphics[width=0.8\textwidth]{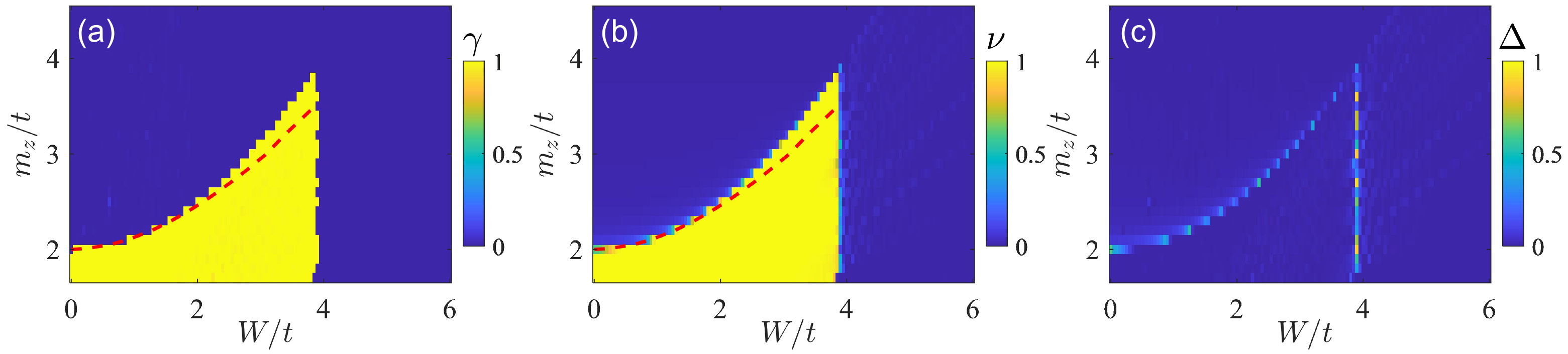}
	\caption{(Color online) (a) Berry phase $\gamma$, (b) winding number $\nu$, and (c) the difference $\Delta=|\gamma-\nu|$ as functions of $W$ and $m_z$. Other parameters are the same as Fig. 2(b) in the main text. The red dash lines in (a) and (b) are the same SCBA calculated phase boundary plotted in Fig. 2(b). Berry phase and winding number are calculated by ED under single particle picture for $L=610$ systems.}
	\label{figS9}
\end{figure}

\subsection{Relation between the quantized Berry phase and winding number in the single-particle case}

As pinpointed in the main text, our model recovers to the 1D AIII class of topological insulators protected by the chiral symmetry in the $W=U=0$ limit. In this single-particle and clean case, the translational invariance of the system preserves. The corresponding Bloch Hamiltonian in momentum $k$ space is $H_0(k)=(m_z-2t\cos k)\sigma_z-2t\sin k\sigma_y$, which satisfies $CH_0(k)C^{-1}=-H_0(k)$ with the chiral symmetry operator $C=\sigma_x$ acting on the two spins (legs). The eigenvalues of $C$ are $\pm 1$ for two eigenstates $|u_\pm(k)\rangle$ satisfying $C|u_\pm(k)\rangle=\pm |u_\pm(k)\rangle$ and $C H_0(k)|u_\pm(k)\rangle=\mp H_0(k)|u_\pm(k)\rangle$. Thus, in the basis of $\hat{S}$ eigenstates, the Bloch Hamiltonian takes an off-block-diagonal form as
\begin{equation}
    H_0(k)=\left(\begin{array}{cc}0& q(k)\\ q^\dagger(k)&0\end{array}\right),
\end{equation}
with $q(k)=2te^{ik}-im_z$. The topological properties in this case can be characterized by the 1D winding number \cite{Mondragon-Shem2014}:
\begin{equation}
    \nu = \frac{1}{2\pi i}\int_0^{2\pi} dk\frac{\partial_k q(k)}{q(k)},
\end{equation}
with $\nu=1$ for $|m_z/t|<2$ and $\nu=0$ for $|m_z/t|>2$. Alternatively, the band topology can be characterized by the quantized Berry phase (in units of $\pi$):
\begin{equation}
    \gamma = \frac{1}{\pi}\int_0^{2\pi} dk \langle u_+(k)|i\partial_k|  u_+(k)\rangle ~~ \text{mod} ~2.
\end{equation}
One can easily verify the relation $\gamma=\nu$ for this clean chiral AIII system with the translation symmetry.

In the presence of disorders, the translational symmetry of the system is broken and $k$ is no longer a good quantum number. In this case, the Berry phase (for a given disorder configuration) can be computed under twisted periodic boundary condition at half filling:
\begin{equation}
	\gamma=\sum_{l=1}^{L} \frac{1}{\pi} \oint d\theta \bra{\psi_l(\theta)} i\partial_{\theta} \ket{\psi_l(\theta)}~~ \text{mod} ~2,
\end{equation}
where $|\psi_l(\theta)\rangle$ is the $l$-th single-particle wave function with the twisted phase $\theta$ obtained from the ED. The numerical results of the Berry phase (averaged over disorder configurations) as functions of $W$ and $m_z$ for $L=610$ systems are shown in Fig.~\ref{figS9}(a) [also in Fig. 2(b) in the main text]. One can use an alternative approach to numerically compute the winding number of disordered systems, based on a covariant real-space generalization of $\nu$ to AIII systems without translational symmetry~\cite{Mondragon-Shem2014}:
\begin{equation}
\nu=\frac{1}{2L'}\mathrm{Tr}'(SP[P,X]),
\end{equation}
where $P=\sum_l(\ket{\psi_l}\bra{\psi_l}-C\ket{\psi_l}\bra{\psi_l}C^{-1})$ sums over the lowest half energies of the single particle wave functions $\ket{\psi_l}$, $L'$ is the number of states summed in $P$, $X$ is the coordinate operator with entry $X_{js,j's'}=j\delta_{jj'}\delta_{ss'}$ ($j$ for lattice site index and $s$ for spin index), $C=\sigma_1^x\otimes\cdots\otimes\sigma_L^x$ is the chiral operator in the real space, and $\mathrm{Tr}'$ represents the trace over the interval with length $L'$. Figure~\ref{figS9}(b) shows the numerical results of the real-space winding number $\nu$ on the $W$-$m_z$ plane from the ED for $L=610$. In Fig.~\ref{figS9}(c), we plot the difference $\Delta=|\gamma-\nu|$ between the quantized Berry phase and real-space winding number as functions of $W$ and $m_z$. One can clearly see that the above two numerical approaches are effectively characterizing the same topology with $\Delta\approx0$, except the region near the topological phase boundary where the real-space winding number is significantly affected by the finite system size.

\end{document}